# An intelligent routing approach using genetic algorithms for quality graded network

## T.R. Gopalakrishnan Nair

Saudi Aramco Endowed Chair,
Technology and Information Management,
Prince Mohammad University,
Kingdom of Saudi Arabia
and
Dayananda Sagar Institutions,
Bangalore, India
Email: trgnair@ieee.org

## Kavitha Sooda*

Nitte Meenakshi Institute of Technology,
P.B. No.6429, Gollahalli, Govindapura,
Yelahanka, Banglore, India
Email: kavithasooda@gmail.com
*Corresponding author

**Abstract:** Intelligent routing in networks has opened up many challenges in modelling and methods, over the past decade. Many techniques do exist for routing on such an environment where path determination was carried out by advertisement, position and near-optimum node selection schemes. In this paper, an efficient routing scheme has been proposed using genetic algorithm for a grade-based two-level node selection method. This method assumes that nodes have the knowledge of its environment and is capable of taking decision for route discovery. The data learnt from the topology which is under consideration for routing, is saved in its local memory. In this two-level node selection scheme, the route discovery operation takes place in multiple levels. At the first level, the grade based selection is applied for considering the most optimal nodes which would be fit for sending data. At the second level, the optimal path is discovered using Genetic Algorithm. The simulation result shows that faster convergence of path took place in the case of the proposed method with good fitness value, as compared to non-graded network.

**Biographical notes:** T.R. Gopalakrishnan Nair holds MTech (IISc, Bangalore) and PhD degree in Computer Science. He has 3 decades experience in Computer Science and Engineering through research, industry and education. He has published several papers and holds patents in multi domains. He is winner of PARAM Award for technology innovation. Currently he is the Saudi

*T.R.G. Nair and K. Sooda*

Aramco Endowed Chair, Technology and Information Management, PMU, KSA and VP of Research and Industry in Dayananda Sagar Institutions, Bangalore, (Sabbatical), India.

Kavitha Sooda holds MTech in Computer Science and Engineering. She has ten years of teaching experience and pursuing her PhD from JNTU, Hyderabad. Her interest includes routing techniques, QoS application, cognitive networks and evolutionary algorithms. Currently she works as Assistant Professor at Nitte Meenakshi Institute of Technology, Bangalore, India.

# 1 Introduction

One of the fast changing areas in internetwork routing is the application of intelligence into it. Among the major challenges that exist in network, the coexistence of multiple infrastructure systems and different models are some of the prime issues. The existing infrastructure lacks the intelligent awareness and coordination of numerous components and applications running on the network and it has opened up various complexities. Also, the coordination with different models needs a good understanding of the services provided by the layers. These two criteria had led the researchers to think of making the network nodes more intelligent. It is reasonable to say that intelligence in the nodes must include properties like learning, reasoning and decision making. To establish accurate results for optimal path, a network must be made aware of its environment. This can be realised by helping the network to learn, think and remember and hence making it intelligent.

The intelligence aspects in network have been dealt in cognitive networks (Fortuna and Moharcic, 2009; Mitola, 2001), autonomic networking (Raymer et al., 2008) and by evolutionary algorithm approach (Dasgupta, 2006). Certain aspects of routing like the energy awareness, QoS, multicasting and network management has been discussed by heuristic algorithms, bio-inspired algorithms and human immune system approaches. These methods are currently applied for making the network intelligent.

This paper presents the implementation of an intelligent system which can carry out routing of packets by taking decision at node level. This makes the routing scheme more efficient and better as decisions are taken twice, i.e., once by Level-1 operation and next by Level-2 operation. Results show that the grading approach along with genetic algorithm (GA) has achieved the required routing path with promising speed.

The rest of this paper is organised as follows: Section 2 addresses the routing process. Section 3 gives the related work. The techniques dealing with fitness function, crossover and mutation methods are shown in Section 4. The grade selection approach is dealt in Section 5. The simulation results are discussed in section 6 and the conclusions and future works are dealt with in section 7.

# 2 Routing

In aiding the path selection in networks, routing algorithms play an important role. A good routing algorithm should be able to find an optimal path and must be simple. It must have low overhead, be robust and stable. Also it needs to converge rapidly, and must remain flexible. There exist a lot of routing algorithms which have been developed for specific kind of networks as well as for general routing purposes. The existing algorithms are either, table driven or demand-driven protocols. These algorithms have been used for different applications depending on their specifications. Still, dynamics of nodes, hidden terminals, power aware routing and location-aid routing remains a challenge in addition to the QoS and multicasting issues. With all these limitations, the protocols have been supporting the existing users of the internet. But the current trend in the usage of the internet shows that these protocols cannot be efficient and self-sustainable for meeting the exploding requirement of billion scale end-to-end delivery.

This would require a method to handle networks in autonomic modes with enormous cognitions for management and decision making capability. Such systems must have enough knowledge base reference where the node would learn, infer and anticipate scenarios and solutions and make use of recollections whether it has come across such scenarios previously. This requires the network to be composed of elements that through learning and reasoning, dynamically adapt to varying network conditions in order to optimise end-to-end performance. This will lead to meeting the requirements of the network as a whole, rather than the individual network components as discussed by (Nair et al., 2008). This is where cognitive networks play an important role. Here cognition is used to observe the forward channels and making behavioural adjustments to seek the best path. It receives the feedback from other nodes while learning which is based on the QoS parameters. The elements of this cognitive network are capable of assembling and incorporating the information from surroundings. It helps them to predict the forward behaviour of the network based on the current states. The performance parameters observed in a network node are collected and uploaded into the network by each cognitive element for decision-making. The decision-making process uses reasoning to determine the next set of actions that can be implemented in the network. Learning and predicting operation in a router depends upon the QoS parameters considered. For this purpose, all the router's forward conditions are sent to its neighbouring routers. Information is shared among routers by sending Cognitive Packets as in (Gelenbe, 2001). These cognitive packets demands negligible channel bandwidth compared to the regular packets carrying information, thus minimising the demand of capacity for cognition activity in the network. At this point when the user wishes to forward the data to a particular end user the information about other nodes connected from the sender to the rest of the participating nodes will be collected by the intelligent nodes. These sets of nodes participate in the selection process of graphical node representation. If the nodes are selected then GA is applied for path selection. From this the best path will be selected based on the shortest hop count.

## 3 Related works

The phenomenon of ever increasing number of users of the Internet has opened up the challenges of providing effective connectivity along with sufficient bandwidth to individual users. The current routing algorithms mainly classify themselves to either static or dynamic routing. In static routing approach routing decisions is merely based on table information at the node level where as in dynamic routing decision is based on only the information that has been collected on few queries requested. Both the approaches do not have the environment awareness knowledge. This is where intelligent routing takes it role and few players like IBM and Motorola have been working on it. The knowledge about the environment has been derived from concepts involving brain theory, evolutionary algorithms, game theory, fuzzy logic and artificial intelligence.

**Figure 1** Cognitive cycle – learning process. The goals are always adhered to, while learning

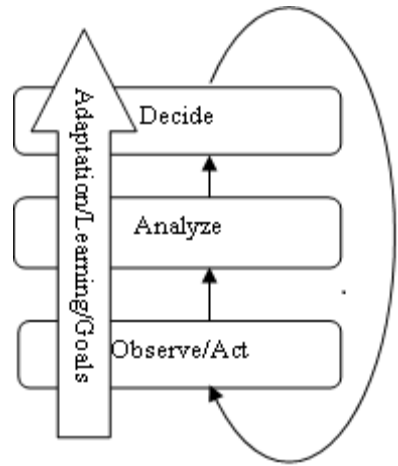

The cognitive cycle as illustrated in Figure 1, was first derived by Mitola (John, 1986), and later the idea of feedback was derived from it. Cognition approach application became acceptable in many fields and also in network. Mitola showed that for any system to become self sustainable it will need to satisfy self-* properties. The challenge in forming such a system is the knowledge representation for the working patterns of it. Understanding the system capability, the functional requirements of the user and mapping of the requirement to the system design along with the limitation of the system itself require a lot of in-depth surveys of the system (Nair et al., 2008).

In order to make the present network systems to be intelligent there is a need for an open platform for cognitive experiments. A common building block has been proposed in (DaSilva et al., 2008). Bandwidth availability has been determined by multi hop analysis (Xiliang et al., 2008). Setting up the geographical layout for cognitive networks is described in (Lokuge et al., 2009). A model which combines a reconfigurable core and control systems along with genetic algorithms for cognitive functionality has been dealt in (Nolan et al., 2007). Genetic algorithm for route discovery has been proposed in (Gelenbe, 2006). The security aspects are dealt in (Prasad, 2008) which discuss the research challenges for security in cognitive networks. Among the major key security aspects which are dealt in (Prasad, 2008), the communication control channel jamming congestion is automatically avoided by our approach a data is forwarded based on the availability of bandwidth at the given link.

The intention objective of this paper is to find an efficient solution for end-to-end delivery which involves geographical intelligence and multiple router integration at large distance. However we handle several layers of routers to prove GA based selection of channel which can be used in graded network routing.

## 4   Genetic algorithms

Genetic algorithms (GA) are a part of evolutionary computing. It is also an efficient search method that could be used for path selection in networks. These stochastic search algorithms are based on the principle of natural selection and recombination. GA (Chang and Ramakrishna, 2002) has been an efficient search methods based on principles of natural selection and genetics. They are being applied successfully to find acceptable solutions to problems in business, engineering, and science. We can find good solution effectively if adequate amount of data is at hand, but as the complexity of data increases GA takes more time to find the solution. GA works well for network model to find the optimal path. In this, the source and the destination nodes are sure to participate in every generation. Other nodes or the genes become a part of the chromosome if they find an optimal path between the source and destination.

A GA is composed with a set of solutions, which represents the chromosomes. This composed set is referred to population. Population consists of set of chromosome which is assumed to give solutions. From this population, we randomly choose the first generation from which solutions are obtained. These solutions become a part of the next generation. Within the population, the chromosomes are tested to see whether they give a valid solution. This testing operation is nothing but the fitness functions which are applied on the chromosome. Operations like selection, crossover and mutation are applied on the selected chromosome to obtain the progeny. Again fitness function is applied to these progeny to test for its fitness. Most fit progeny chromosome will be the participants in the next generation. The best sets of solution are obtained using heuristic search techniques. The general description of GA is as follows:

a   **First Generation** randomly pick n chromosome to form a population assuming that this could be the probable solution to the problem.

b   **Fitness Function** the fitness function f(x) is applied on each chromosome in the generation.

c   **Next Generation** create the next generation by performing the following steps until n chromosomes are obtained,

   i.   **Selection operation** Select any two best fit chromosomes from the generation.

   ii.  **Crossover** with a defined probability. Apply the crossover technique for the above obtained chromosome to form the children.

   iii. **Mutation** with a defined probability. Mutate a new gene at desired position.

d   **Test** whether the obtained children are fit to go to next generation. If yes, then move them to next generation.

e   **Test** if generation is of desired size, if yes, stop, and return the best solution from the current generation.

f   **Repeat** go back to b

The performance of GA is based on efficient representation, evaluation of fitness function and other parameters like size of population, rate of crossover, mutation and the strength of selection. Genetic algorithms are able to find out optimal or near optimal solution depending on the selection function as described by Goldberg (1995) and (Ray et al., 2004).

The fitness function (Nair and Sooda, 2010) used for the selection of the chromosome is as follows:

$$f_j(t) = \frac{B_j(t)}{\sum_{i=0}^{l} B_i(t)} \qquad (1)$$

### 4.1 Natural selection

GA uses a selection mechanism to select the individuals from the population to insert into the mating pool. Individuals from the mating pool are used to generate new off spring which will participate in the next generation. As the individuals of the next generation are going to participate further this is better for the genes to be of good condition. The selection mechanism is a process of selecting the individuals of good condition. This selection function leads to a better population with good condition. The convergence rate largely depends on the selection function.

Here, the selection process is carried out by Roulette Wheel method as shown in Figure 2. In this method, the individuals are chosen based on the relative fitness with its competitors. Here the wheel is divided into slice, where the fit chromosome gets larger slice. For selecting the chromosome for next generation the wheel is spun. Once the wheel stops, the individual corresponding to the slice on which it lands goes to next generation.

**Figure 2** Roulette wheel

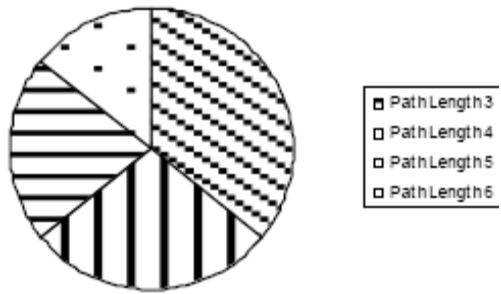

### 4.2 Singlepoint/multipoint crossover

Crossover operator (Murata and Ishibuchi, 1996) combines sub parts of two parent chromosomes and produces off spring that contains some parts of both the parent. Here we consider both single point and multipoint crossover technique. In single point crossover technique, one offspring consists of first part of one parent and second part of the other parent. Similarly the other offspring is generated. Here, we also use multi point crossover mechanism called partially mapped crossover. In this, two chromosomes are picked at random. The strings between the crossover points are exchanged position by position; other elements are determined by ordering information, which is partially determined by each of its parents.

For example:

Single point crossover:

Parent 1: 1 2 3 4 | 5 6 7

Parent 2: 1 3 4 6 | 2 5 7

Offspring 1: 1 2 3 4 2 5 7

Offspring 2: 1 3 4 6 5 6 7

Multipoint crossover:

Parent 1: 1 2 3 | 4 5 | 6 7

Parent 2: 1 3 4 | 6 2 | 5 7

Offspring 1: 1 3 5 6 2 4 7

Offspring 2: 1 2 6 4 5 3 7

Here the simulation results are obtained based on the nodes which participate in the path determination. The various paths from source to destination are considered to be the chromosomes. We have fixed the position for single point crossover as four and for multipoint crossover as 3 and 5. These are the gene position of the path considered. Thus the offspring are derived according to the explanation given above.

### 4.3 Mutation

Sometimes it may be possible that by crossover operation, a new population never gets generated. To overcome this limitation, we do mutation operation. Here we use insertion method (Lee and Lim, 2005) as a node along the optimal path that may be eliminated through crossover.

## 5 Grade selection approach

Grade value estimation method for implementing intelligent routing (Nair and Sooda, 2010) in the autonomic network is the core focus of research thrust. Grade is like an index; it is made available everywhere and routing will much depend on it. It signifies the quality of the router, which is in fact the knowledge of the environment. The router must be an intelligent entity because it performs different operations based on the information from environment. It depends on input, output, load and resource availability.

### 5.1 Necessity of grading

Grade largely provides information about the path. All kinds of knowledge about the path make the routing efficient. Then the efficiency depends on the kind of knowledge base that is possessed by the node. Many researchers have defined grade as hop distance (Tong et al., 2010) and have obtained better results by considering the value of the grade. As there is a lot of external influence on the node in the autonomic network, the network topology is dynamic in nature. Hence the condition of all nodes must be made available everywhere. This defines the health condition and utility factor of that particular node. Overall network grade in local vicinity can be done by any heuristic algorithm, which shall search and find the best possible nodes that can participate in routing. Grading and producing graded parametric surface in network depends on the intelligent level chosen for operation. When varying the level of intelligence is considered, defining rules and regulation protocol is challenging. Once these factors are well-defined, multi- grade network can be easily achieved.

A grade value must be calculated in real-time based on observed factors. The challenge lies in identifying the right indicator for the input of the grade based function. The choice of the indicator depends on various factors like distributed environment, reliability in the presence of external perturbations, internal perturbations and resource availability. One way to solve this issue is to study the topological network and simply work on pre-defined training cases. Research has proved that a run-time stochastic approach has better results in obtaining the training set. If any deviation is detected, there is an appropriate action taken to overcome it. However, learning happens there at system start-up, which need not be suitable for the application wherein the population is generated randomly later. The challenge also lies in the protocol, which depends on the nodes located at different parts. Collecting information cumulatively at one location to find the best possible indicator is a difficult task. A lot of research scope is yet to be resolved in this direction. A few topics have been dealt with in (Gonzalez-Valenzuela et al., 2008).

Game theory, probability, linear programming, evolutionary algorithm, genetic algorithm, artificial immune system, artificial intelligence and many more stochastic approaches have been applied to achieve the knowledge accumulation from the network. This work is based on the grade function, which takes varying parameters depending on the environment condition. Initially, the work can be carried out for a fixed parameter to understand the working principle of the grade function and to test the efficiency. The grade function can be selected based on:

- elasticity of application traffic
- human psychology based on mean opinion score
- resource allocation efficiency
- fairness of different shapes of utility function, which leads to optimal resource allocation
- rate, reliability, delay, jitter, power level
- congestion level, energy efficiency, network lifetime, collective estimation error.

## 5.2 *Queueing systems*

A queueing model is used for mathematically analysing the queueing behaviour. We can obtain many steady state performance measures, which include: average number in the queue, probability of finding the system in a particular state and a queue being full or empty and statistical distribution of the number of queues. These performance measures are important to be considered with respect to the service offered by the link. Analysis of such a queueing model will help us to identify the issue and the impact of the changes to be assessed. Queueing model is a stochastic model that represents the probability that a queueing system will be found in a particular configuration. It is represented using Kendall's notation. The model used for the current work is M/M/1 (Leonard, 1976). It stands for Markovian inter-arrival time along with exponential service time with a single server. This model represents the steady state of the system under consideration. The exponential distribution time describes the event's occurrence and independence at a constant average rate. The Poisson statistical model is a generally accepted tool to predict end user behaviour. Little's formula for M/M/1 queue for mean number of jobs in the system is given by,

$$E(n) = \frac{\rho}{1-\rho} * \frac{1}{\lambda} \tag{2}$$

Here $\rho$ represents the traffic intensity and $\lambda$ represents the mean arrival rate of the message flow.

### 5.3 Implementation of grade

This paper is based on the following list of parameters which helps in grading according to priority derived from (Nair and Sooda, 2011)

- network lifetime (NL)
- node density (ND)
- traffic congestion (TC)
- resource allocation (RA)
- delay of the packet arrival
- bandwidth availability

### 5.4 Level-1 and Level-2 operations

*Level-1* is applied region-wise and the goal is to achieve favorable routing based on selected attributes. The values obtained from *Level-1* must be able to eliminate the non-production node. Non-production nodes are those that come in the pitfall of congestion and possess less resource availability. These nodes must be identified by the algorithm that defines the gradient from the most non-productive to productive nodes in a homogeneous network. This is identified by assigning a grade value from –3 to +3, i.e. it signifies the productivity value of the node. At this point, we are able to calibrate the routing process region-wise. The algorithm requires proactive decision making on the output obtained. This is because many paths exist to reach the destination, and we have to choose the most optimal path. Once the gradient value has been calculated, it can be made available as pervasive information packets to all the other nodes to obtain the optimal path constitution.

Now the graded function, i.e., *Level-2*, considers the calculated values as its input of *level-1*. The output of this function defines the route availability for the set of nodes considered. This shall be calculated for all the available paths leading towards the destination node. The mean value of the gradient in the grade function shows the success level of operation of the network.

## 5.5 Level-1 operation

Assume top three attributes are selected for every region based on priority assigned (Figure 3).

*Step 1*: The top three priority nodes are selected.

*Step 2*: Select the nodes which are nearing to the optimal (relaxation of ± 2).

*Step 3*: Selection is done for the nodes that would satisfy the relaxation range from 0 to 2.

*Step 4*: Build a connectivity graph for making sure connectivity exists between regions and apply second level grade to find the optimal path (Figure 4).

Assessing the utility parameter increases the grade value. This approach helps in considering the output for routing.

**Figure 3**  Scale of prioritisation assigned to the nodes after observation. Here –3 represents no network lifetime, 0 represents most optimal node and +3 represents nodes which can be considered

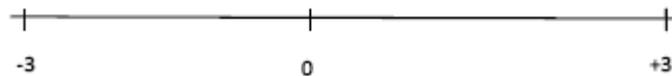

**Figure 4**  Region based network topology

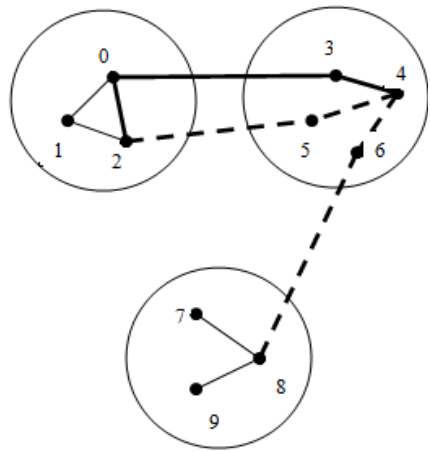

## 5.6 Level-2 operation

The topology obtained from *Level-1* operation is considered as input data here. GA selection mechanism is applied based on the bandwidth availability for the path determination.

*Step 1*: Consider all possible paths from source to destination as input set.

*Step 2*: Execute the following steps for K generations:

a. Select one chromosome based on elitism technique.

b. Select the remaining of the chromosome based on 90% crossover and 10% mutation.

c. Generate N such chromosomes for each generation.

d. Calculate the fitness value of each one of the chromosomes. The chromosome whose fitness value is above 0.9 is carried forward to the next generation.

*Step 3*: Select the chromosome whose fitness value is above 0.9.

Here the fitness is calculated based on the bandwidth available at the node.

### 5.7 Design aspects

The design involves the generation of an input model, priority model, gradient algorithm and knowledge base. The input model is based on the M/M/1 technique. The priority model is obtained as shown in Figure 4. The gradient algorithm is the *Level-1* operation as discussed earlier. The knowledge base contains the nodes and paths with good condition. Six parameters are considered to obtain the required result. The following parameters are made available in a vector:

a. Resource allocated (RA)

b. Network lifetime (NL)

c. Bandwidth at nodes

Figure 5 depicts the input M/M/1 model after which we next apply the priority model from which certain number of nodes gets selected by *Level-1*. Next we apply *Level-2 operation*, which is the gradient algorithm by which we determine the optimal path. This is the knowledge gained by the network and it is fed to the knowledge base.

**Figure 5** Flow diagram of the implementation

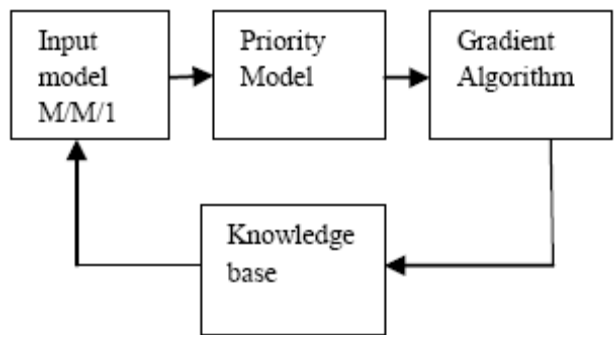

*Delay*, *traffic congestion* and *Node Density* are calculated. *Delay* is derived from service rate, arrival rate and capacity. *Traffic Congestion* (TC) is derived from the expected data rate at the nodes. *Node Density* (ND) to be calculated is based on in-degree of the topology set-up. Information about the connectivity must be made available in a file. For the first level of node selection we apply the first level selection method. The nodes are prioritised based on local observation. Prioritising is performed based on ontological reasoning as shown in Figure 6.

From this top three priority nodes are selected. Checking is carried out to see whether delay has reduced and resource allocation is made available. Verification is also performed for traffic congestion. If all three parameters are satisfied then nodes can participate for the next level of selection.

**Figure 6** Priority model

```
IF (NL) {
    IF (ND < 5) {
        IF (TC does not exist) {
            IF (RA) {
                IF (Delay does not exist) {
                    P=1;
                ELSE
                    P=2;
            ELSE
                P=3;
        ELSE
            P=4
    ELSE
        P=5;
ELSE
    P=6;
```

Now we categorise the favourable and non-favourable nodes and look for connectivity between the regions. Select the source and destination node. The next step involves checking the bandwidth availability and applying GA with quality grading of nodes. These procedures lead to the realisation of optimal path, as discussed earlier.

*5.8 Delay calculation*

We assume that the node location, external traffic requirements $\gamma_{ik}$, channel cost $di(Ci)$, the constants $D$, $\mu$ and the flow ($\lambda i$) are given and feasible. Thus the Delay $T$ at the node is given by,

$$T = \sum_{i=1}^{M} \frac{\lambda i}{\gamma} \left[ \frac{1}{\mu Ci - \lambda i} \right] \quad (3)$$

The average rate of message flow $\lambda_i$, on the *i*-th channel is equal to the sum of the average message flow rate of all paths that traverse this channel. The traffic entering the network from the external sources forms a Poisson's process with a mean $\gamma_{jk}$ (messages per second) for those messages originating at node *j* and destined for node *k*. Therefore, the total external traffic entering and leaving the network are considered to be equal to $\lambda_i$.

*An intelligent routing approach using genetic algorithms*

## 6 Simulation and results

The topology was set up using region-based design approach. A random topology was set-up and tested for different topological structures. Initially, the path selection took place region-wise. Later the regions were considered for connectivity based on the initial setup. Six parameters are considered for the simulation. Five parameters were considered by *Level-1* and the sixth parameter was considered by *Level-2* approach which involves GA. Table 1 describes how the values were obtained.

**Table 1**     Input parameters

| Parameter list | Description |
| --- | --- |
| Bandwidth at nodes | Random number at initial level |
| Congestion | Calculated based on bandwidth available and current traffic |
| Delay | Calculated based on M/M/1 model |
| Network lifetime | Random number |
| Node density | Number of in-degree to the node |
| Resource allocated | Random number |

The GA operations are carried out with the probability values as shown in Table 2, on the graded networks.

**Table 2**     GA properties

| GA Properties | Values chosen |
| --- | --- |
| Crossover Probability | 95% |
| Mutation Probability | 5% |
| Population Size | 4 to 256 nodes |
| Number of generations | 10 |
| Termination Condition | Terminate when destination node is reached with sufficient bandwidth on all possible intermediate paths or number of generation reaches zero |

Now *Level-1* operation is applied to obtain a concise network topology which would have the knowledge of the environment where the topology belongs. Most of the learning phases were carried out at this stage. Here the nodes which have the maximum lifetime, least congestion and nodes with resource availability have been selected. The rest would be considered as non-productive nodes and would not participate for the next level grade selection scheme.

Now the nodes which would have been selected by *Level-1* operation would go as input for *Level-2* operation. Here the bandwidth availability at the node level is considered for the selection. These values were randomly determined and the simulation results are obtained.

Table 3 gives a comparison of optimal path determination of GA with quality grading of nodes and GA without quality grading of nodes for one of the random runs of the model. Here, the GA with quality grading of nodes determines the path based on the quality of the node, thus making it more reliable for best path determination. In GA without quality grading of nodes, the few nodes that were selected were either congested or with less resource, and a few led to congestion. Thus, determining the optimal path took more time which are indicated by low fitness value.

**Table 3** Output analysis

| Total nodes | GA with quality grading of nodes | | | GA without quality grading of nodes | | |
|---|---|---|---|---|---|---|
| | Number of nodes selected | Fitness Value | Route length | Number of nodes selected | Fitness Value | Route length |
| 4 | 2 | .9655 | 2 | 4 | .9655 | 2 |
| 8 | 5 | .9812 | 4 | 6 | .8977 | 7 |
| 16 | 11 | .9411 | 6 | 12 | .9219 | 10 |
| 32 | 24 | .9522 | 9 | 24 | .8111 | 11 |
| 64 | 28 | .8216 | 12 | 33 | .7644 | 15 |
| 128 | 35 | .9911 | 6 | 38 | .8622 | 9 |
| 256 | 34 | .9333 | 9 | 42 | .8151 | 12 |

Table 4 shows that the GA with quality grading of nodes was better in term of route length as the computation was based on state aware network and intelligence aspect with certain QoS parameters consider at level-1 operation. The results show that better and more deterministic paths can be obtained by the method proposed.

**Table 4** Comparison of the two methods

| Issues | GA without quality grading of nodes | GA with quality grading of nodes |
|---|---|---|
| Level-1 parameters | No | Yes |
| Knowledge base | No | Yes |
| Shortest path | 83% | 90% |
| Intelligence aspects | No | Yes |

Figure 7 shows that the number of nodes selected by the GA with quality grading of nodes, was lesser and more reliable as the paths were determined by the quality of the nodes than the GA without quality grading of nodes. The number of nodes that were selected is also listed in Table 2. The lesser nodes obtained by the quality grading signifies the reduction of search space for determining the optimal path. These nodes are more reliable as they have better packet lifetime, less density ahead and therefore occurrence of congestion was less. Simulation results showed that there was 1% chance for congestion to occur once the path was fixed. However, this is negligible as selections of the nodes are made on five parameters at level-1 which would minimise the occurrence of such scenario.

Figure 7 also shows that the performance of GA without quality grading of nodes are lesser or equal to performance of GA with quality grading of nodes. But the chance of congestion in GA without quality grading of nodes were 20% more than GA with quality grading of nodes. The reason is that the node selection was not based on the quality selection scheme. The troughs and valleys of parametric surface of future networks can be made use of effectively in this way leading to autonomic management of intelligent networks. Figure 7 shows the advantage of obtaining less number of nodes as optimal path determination becomes faster in quality grading scheme. With lesser number of nodes been selected in graded network we can update optimal path much faster thereby increasing routing efficiency. The results given here are obtained from 25 runs for different topology obtained with varying parameters listed in Table 1 and for each run the number of nodes was varied from 4 to 256.

**Figure 7** The graph shows the nodes selected by the two methods (see online version for colours)

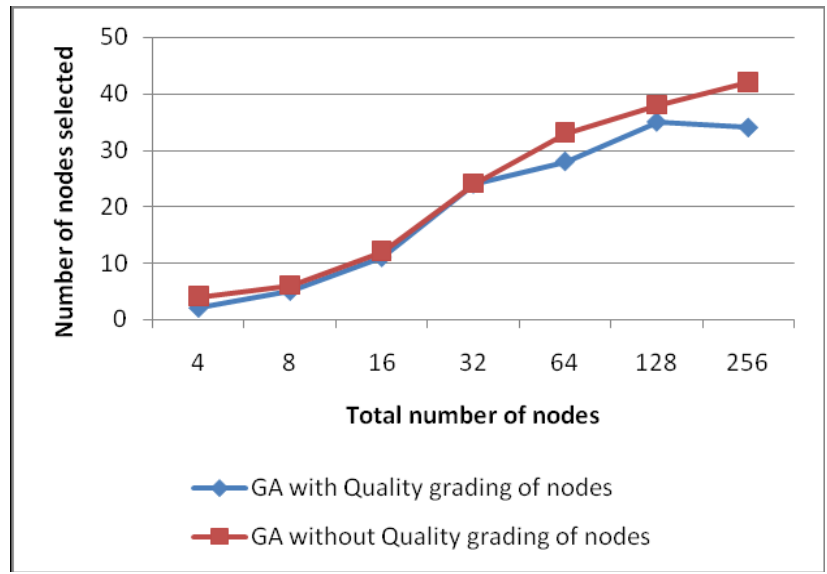

## 7 Conclusions and future work

The results obtained show a significant improvement in the convergence of optimal path by using GA with quality grading of nodes. The comparative results of the two approaches show that modified GA selected only the best nodes for path determination when compared to GA without the awareness of the state of the network emphasising grading effect. Further, the algorithms may be improved with multi-parameters (Dipankar, 2006) that need to be considered to assess the grade function to be more reflective of network states. This can lead to a homogeneous grading or a highly graded network that will be organising more collective information at the nodes, where decision can be derived by intelligent arbitration.